\documentclass[12pt]{article}
\usepackage{amsmath} %
\usepackage{amssymb}
\usepackage{indentfirst}
\usepackage{placeins}
\date{}
\usepackage{natbib}
\usepackage{graphicx} %
\usepackage{xcolor}
\definecolor{mydarkblue}{rgb}{0,0.08,0.45}
\usepackage[colorlinks,citecolor=mydarkblue]{hyperref}
\usepackage[capitalize,noabbrev]{cleveref}
\usepackage{booktabs}
\providecommand{\keywords}[1]{\textbf{Keywords:} #1}
\usepackage{multirow}
\usepackage{enumitem}
\usepackage{csquotes}
\usepackage{circledsteps} 

\usepackage[table]{xcolor}
\usepackage{longtable}

\begin{document}
\title{Making Power Explicable in AI: \\
Analyzing, Understanding, and Redirecting Power to Operationalize Ethics in AI Technical Practice}
\author{Weina Jin, Elise Li Zheng, Ghassan Hamarneh
}
\maketitle

\begin{abstract}

The operationalization of ethics in the technical practices of artificial intelligence (AI) is facing significant challenges. To address the problem of ineffective implementation of AI ethics, we present our diagnosis, analysis, and interventional recommendations from a unique perspective of the real-world implementation of AI ethics through explainable AI (XAI) techniques. We first describe the phenomenon (i.e., the “symptoms”) of ineffective implementation of AI ethics in explainable AI using four empirical cases. From the “symptoms”, we diagnose the root cause (i.e., the “disease”) being the dysfunction and imbalance of power structures in the sociotechnical system of AI. The power structures are dominated by unjust and unchecked power that does not represent the benefits and interests of the public and the most impacted communities, and cannot be countervailed by ethical power. Based on the understanding of power mechanisms, we propose three interventional recommendations to tackle the root cause, including: 1) Making power explicable and checked, 2) Reframing the narratives and assumptions of AI and AI ethics to check unjust power and reflect the values and benefits of the public, and 3) Uniting the efforts of ethical and scientific conduct of AI to encode ethical values as technical standards, norms, and methods, including conducting critical examinations and limitation analyses of AI technical practices. We hope that our diagnosis and interventional recommendations can be a useful input to the AI community and civil society’s ongoing discussion and implementation of ethics in AI for ethical and responsible AI practice.
\end{abstract}

\section{Introduction}\label{sec:intro}

The state-of-the-art data-driven technologies, now claimed under the name of artificial intelligence/machine learning (AI/ML)~\citep{Suchman2023,Goodlad2023}, have gained great momentum and societal attention in the past decade. While AI technologies have the potential to bring positive changes, the negative social impacts they created have already raised concerns, including: algorithmic surveillance~\citep{zuboffAgeSurveillanceCapitalism2019,acemogluPowerProgressOur2023}, discrimination~\citep{chunDiscriminatingDataCorrelation2021,nobleAlgorithmsOppressionHow2018,dignazioDataFeminism2020}, labor exploitation~\citep{macgillisFulfillmentWinningLosing2021}, deprived autonomy, rights, and power of workers and users~\citep{neda_surrogate_2019,Law_2023,gray_ghost_2019}, work degradation~\citep{ResnikoffJason2021,Rosenblat2018-vw}, widened inequality~\citep{eubanksAutomatingInequalityHow2018, Acemoglu2022,oneilWeaponsMathDestruction2016}, misinformation~\citep{MarcusGaryF2024TSV, ethicalMISyn}, privacy breach~\citep{MarcusGaryF2024TSV}, copyright infringement~\citep{GenerativeAIHas,MarcusGaryF2024TSV}, and environmental damage~\citep{crawfordAtlasAIPower2021,li2023makingaithirstyuncovering, JMLR:v21:20-312,10.1145/3442188.3445922}. 

\begin{figure}
    \centering
    \includegraphics[width=\linewidth]{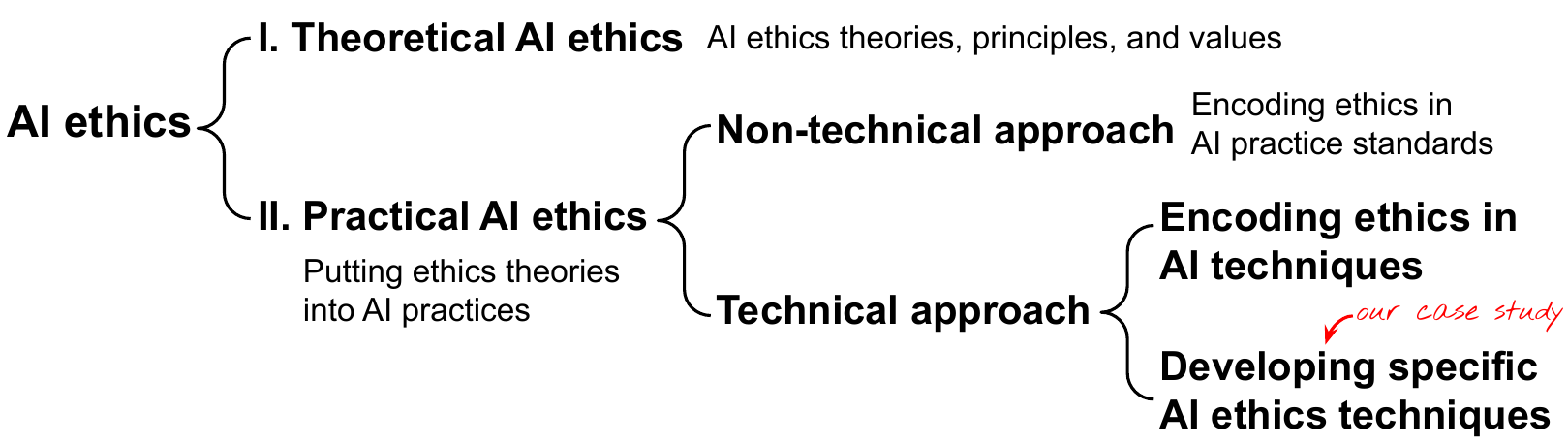}
    \caption{The scope of AI ethics in our paper. The problem of ethics implementation in AI technical practices focuses on II. practical AI ethics that implements I. AI ethics theories, principles, and values into the AI technical professionals’ day-to-day practices and activities in the whole pipeline of AI research, development, deployment and maintenance, etc.
}
    \label{fig:scope_ethics}
\end{figure}

Given the ongoing and anticipated harms and risks of AI, ``\textbf{how to conduct AI research and development ethically and responsibly}'' becomes a pressing problem for both the AI community and the public. AI ethics is an indispensable component to safeguard the responsible development of AI, along with other components such as regulations, policies, and laws. As a subfield of applied ethics, AI ethics is to ensure that AI technologies are researched, designed, developed, and deployed ethically, and the risks and harms of AI to the marginal, vulnerable, or most affected communities and society are minimized. The realm of AI ethics can be roughly categorized into two major aspects: theoretical and practical AI ethics (\cref{fig:scope_ethics}).

Theoretical AI ethics is to theoretically understand the moral aspect of AI technologies regarding what is good and bad. A notable progress in theoretical AI ethics is the development of AI principles, similar to the principles in biomedical ethics~\citep{Floridi2018}. Behind the principles are different ethical theories shaped by various philosophical traditions~\citep{Hagendorff2020,BIRHANE2021100205}. 

Practical AI ethics is to implement ethics in AI technical practices\footnote{In this paper, we use the phrase “AI practices” or “AI technical practices” to denote the day-to-day practices and activities of AI technical professionals in the whole pipeline of AI research, development, deployment and maintenance, etc.}; the implementation of AI ethics can be in non-technical or technical approach: the non-technical approach of AI ethics implementation can include norms, standards, procedures, codes of conduct, and ethics review in AI technical practices; and the technical approach of AI ethics implementation can include two aspects: 1)  translating and encoding ethical principles into the technical design, evaluation, and implementation of AI techniques in various AI problems, tasks, subfields, techniques, and applications; and 2) developing specific techniques to ensure certain ethical properties of AI, such as fairness, transparency, accountability, explainability, privacy, and safety.

Our work focuses on practical AI ethics and the problem of the real-world implementation of ethics into AI technical practice. Despite recent advances of the development of AI ethics principles and fast-growing AI ethics techniques of fairness, explainability, and privacy, 
operationalizing ethics into AI technical practice is facing significant challenges, as summarized in C1-C3:
\begin{enumerate}[leftmargin=*,label=\textbf{C\arabic*}]
    \item \textbf{Abstract ethical principles vs. concrete realities.} There is a significant gap between the theory and practice of AI ethics. The implementation and translation of AI ethics into real-world AI practice is not fully supported~\citep{Bleher2023,Hagendorff2021,Munn2022,Ayling2021,Hagendorff2020,Morley2019,Mittelstadt2019,McLennan2022}. 
    \item \textbf{Ethics for post-hoc fix vs. ethics encoded in the AI blueprint.} 
    Even if AI ethics is included in the technical development, ethics mainly functions at a superficial level~\citep{10.1145/3531146.3533157, greeneBetterNicerClearer2019a,Floridi2019}, usually for post-harm fixes. That means AI ethics techniques (such as bias mitigation and explainable AI) are implemented to improve ethical features or mitigate side effects/harms of AI, but the motivation to develop an AI technique in the first place is rarely questioned, challenged, or prohibited by AI ethics~\citep{doi:10.1177/2053951720942541}. Furthermore, there is relatively less support for the methodologies and technical practices of encoding ethical values from the outset in the AI blueprint, and investigating and preventing ethical risks from the roots~\citep{ai_imagination,Burrell2024Automated,birhane2022values}. 
    \item \label{item: ethics_washing} \textbf{Ethics washing.} 
    During the process of implementing AI ethics, there are unethical risks such as ethics washing\footnote{The definitions of these terms are provided in~\cref{app:glossary} Glossary.}, ethics shopping, using ethics merely for branding and marketing, and using ethics as self-regulation to escape regulations~\citep{Floridi2019, Wagner2019, 10.1145/3351095.3372860,katzArtificialWhitenessPolitics2020, greeneBetterNicerClearer2019a,schaakeTechCoupHow2024,doi:10.1177/2053951720942541,Ochigame2022}. In these circumstances, ``performing ethics''~\citep{Greenhalgh2024} itself becomes the end goal, not the means to an end of conducting AI practices ethically and responsibly~\citep{10.1145/3351095.3372860}. 
\end{enumerate}

\begin{table}[!t]
    \centering
    \caption{The power-sensitive interventional recommendations for ethical and responsible AI technical practices proposed in~\cref{sec:framework}. }
    {\renewcommand{\arraystretch}{1.2} 
    \begin{tabular}{p{0.01\linewidth}p{0.9\linewidth}}
    \bottomrule
\rowcolor[gray]{.9}  
\textbf{1.} & \textbf{Making power visible, explicable, and checked} \\ \toprule
       $\square$  & Embed power analysis in AI technical practices, i.e., asking “for whom?” and “who can be put at risk?” in major and minor design choices in AI \\ \hline
        $\square$  & Embed power analysis in the routine of ethics review and ethics analysis of AI techniques \\ \hline
         $\square$  & Unite just power from the AI community and civil society to take collective actions to check and balance the current dominant power in AI \\[1.8em]
    \bottomrule
\rowcolor[gray]{.9}  
\textbf{2.} & \textbf{Reframing AI and AI ethics for justice} \\ \toprule

$\square$  & Reframe AI developmental goals to include the mitigation of harms, in addition to technical innovation (analogous to building a good car, not a good engine only)\\ \hline
$\square$  & Reframe AI ethics as the necessary adversarial guardrail for the healthy development of AI (analogous to regarding the adversarial role of the brake not as a tradeoff with the engine)\\[1.8em]
    \bottomrule
\rowcolor[gray]{.9}  
\textbf{3.} & \textbf{Encoding ethics as technical and scientific methodologies in AI practice } \\ \toprule
$\square$  & Critically reflect and examine the taken-for-granted practices and their assumptions in AI, and institutionalize critical examination as a major AI research agenda\\ \hline

$\square$  & Investigate in equivalent amount of resources and attention to algorithmic limitation analysis as the attention on algorithmic performance analysis, and institutionalize limitation analysis as the AI technical evaluation standard \\ \hline

$\square$  & Explicitly encode ethical values and perspectives as technical standards for an AI task or subfield \\ 
\bottomrule
    \end{tabular}}
    \label{tab:recommand}
\end{table}

This work aims to address these challenges by revealing and tackling an important latent factor in these roadblocks: power. Previous works have advocated for the lens of power in AI ethical analysis to complement the existing principled approach and to release the practical and emancipatory potential of AI ethics~\citep{doi:10.1080/15265161.2024.2377139,Waelen2022, BoenigLiptsin2022,Burrell2024,10.1145/3531146.3533157,Kalluri2020}. 
Our work joins these pioneering efforts and provides a unique perspective from our research practice in the technical AI ethics of explainable AI (XAI).
We first describe the phenomenon (i.e., the “symptoms”) of ineffective implementation of AI ethics in XAI in four empirical cases (\cref{sec:case}). Based on the “symptoms”, we diagnose the root causes (i.e., the “disease”) of power being the latent factor that can interfere with the actual operationalization of ethical AI practice (\cref{sec:analyze}). We also analyze how power works in AI practice by drawing on theories from social science and epistemology. Based on the understanding of the power mechanism, we propose the interventional recommendations to tackle the key points in power mechanism, shown in~\cref{tab:recommand}, including: 1) Making power visible and scrutinizable; 2) Reframing AI and AI ethics for justice; and 3) Encoding ethics as technical and scientific methodologies in AI practice (\cref{sec:framework}). 
Our work contributes to the ethical and responsible AI practice by providing a thorough diagnosis, understanding, and intervention grounded in the real-world implementation experience and phenomenon of AI ethics. We hope our work can be a useful input to the AI community and civil society’s ongoing discussion and implementation of ethics in AI practice.

\section{Technical practice of AI ethics: Four cases of explainable AI}\label{sec:case}

We present four cases in practical AI ethics regarding developing AI ethics techniques to implement ethics in technical practice. The cases are drawn from the authors' (WJ, GH) first-hand research practice\footnote{Unique in our research approach is that our research was carried out in close collaboration with doctors. We also have domain knowledge in medicine (the author WJ holds an MD degree), and have skills in human-centered research methodologies, including qualitative research. } and findings in the field of technical AI ethics on explainable AI~\citep{JIN2023102684,JIN2024102751,jin2024plausibilitysurprisinglyproblematicxai}. Our primary research focus is on explainable AI for medical image analysis. Interpretable or explainable AI (XAI) is usually regarded as a branch of technical AI ethics that studies how to explain AI models in human-understandable ways~\citep{doshi2017towards}. 
XAI is a sentinel field to inspect   the operationalization of ethics in AI technical practices, because explainability is one of the five AI ethics principles~\citep{Floridi2018}, and the XAI field studies how to achieve explainability via AI techniques. The focus of these cases emerged from the pipeline of ethical AI techniques, from the scope, design, evaluation, to technical objective.

\subsection{Case 1: XAI in AI ethics and ``explain to whom?''}
XAI techniques may be designed for different purposes and audiences. In the context of AI ethics, explainability or explicability is one of the five AI ethics principles that ``enabl[es] the other principles through intelligibility and accountability''~\citep{Floridi2018} (also see~\citet{Morley2019,Jobin2019} about the meta-review of AI ethics principles). 
Given that the main motivation of XAI is algorithmic transparency and accountability to enable oversight, 
the main audiences of explainability in AI ethics should be non-technical users and stakeholders\footnote{In the context of AI ethics, despite the fact that technical audiences are important in overseeing AI, the technical development of XAI should set non-technical audiences as the main audience and their purposes as the main objectives of XAI. This is because: 1) The main scenarios when XAI is needed are in human-AI collaborative tasks, usually in high-stakes domains, such as healthcare, criminal justice, finance. Non-technical audiences are the major group of audiences who require XAI for transparency and accountability. 2) In scenarios when XAI is required, technical audiences are more likely to have conflicts of interests with the technical sectors that receive oversight. Thus, developing XAI mainly for technical audiences may face the challenge of “using ethics as self-regulation to escape from regulations” mentioned in \ref{item: ethics_washing} in the Introduction. 3) XAI focusing on technical audiences will create barriers for non-technical audiences to understand the AI explanation. This is a form of epistemic injustice~\citep{Fricker2007} that is unethical. }. We call them ``users" for short throughout this paper. However, in real-world technical practice of XAI research and development, the target audiences of XAI are usually vague, and non-technical audiences are not prioritized. This phenomenon has been observed by multiple researchers and studies, as summarized below.

In the paper \textit{Explainable AI: Beware of Inmates Running the Asylum}, \citet{DBLP:journals/corr/abs-1712-00547} argues that ``most of us as AI researchers are building explanatory agents for ourselves, rather than for the intended users.'' 
This corresponds to the findings of a user study on XAI deployment in the real world~\citep{10.1145/3351095.3375624}: 
“the majority of deployments are not for end users affected by the model but rather for machine learning engineers, who use explainability to debug the model itself. There is thus a gap between explainability in practice and the goal of transparency, since explanations primarily serve internal stakeholders rather than external ones.”
Similarly, in a user study with 12 practitioners working on AI for Social Good (AI4SG) projects  in the Global South~\citep{10.1145/3689904.3694707}, it shows that “[m]any participants who used XAI methods in practice stated that their primary motivation for using XAI was to aid in publishing their papers rather than improving user comprehension. 
...We find that current approaches to using XAI in AI4SG create a severe incentive misalignment, devaluing the needs of users in favor of publication practices in machine learning that often prioritize model performance over user interpretability.”
In \textit{On Two XAI Cultures}, \citet{jiang2021xaiculturescasestudy} commented that “[m]uch of XAI research build explanations for those \textit{who are already AI experts}. Overall, little regard is given to non-AI-expert stakeholders who need AI explanations, and have different goals for those explanations. This produces a large gap between technical experts’ version of XAI, and what non-technical audiences expect of XAI.”

The ambiguity of XAI audiences propagates to the design, evaluation, and purposes of XAI, as detailed in the following cases. 

\subsection{Case 2: XAI technical design on understandability}

\citet{doshi2017towards}’s definition of XAI in the beginning of \cref{sec:case} shows that understandability is the basic requirement in XAI technical design to enable users’ understanding of AI prediction. Like explainability, understandability is always coupled with its audience to whom and in which context an AI explanation is understandable. An XAI technique designed for the technical or “general” audiences and purposes may not be understandable for non-technical audiences in specific contexts\footnote{Examples of XAI for technical audiences’ understandability include “neuron visualization” \citep{Olah2017,bau2020units} and “mechanistic interpretability” \citep{wiegreffe2024mechanistic}. These approaches usually aim to reveal low-level, human-interpretable technical details inside the AI model for purposes such as model debugging, monitoring, discovery, and improvement. These approaches are analogous to neuroscience that probes into the neuron activities in the brain to understand how the brain functions. Technical knowledge of how the AI model is built is usually required to interpret information from these approaches.}~\citep{euca}. Nonetheless, in our research practice in XAI, we found that the research exploration in the technical design of understandability usually stays at a superficial level, and cannot meet users’ specific understandability requirements in concrete contexts. 

\begin{figure}[!h]
    \centering
    \includegraphics[width=0.65\linewidth]{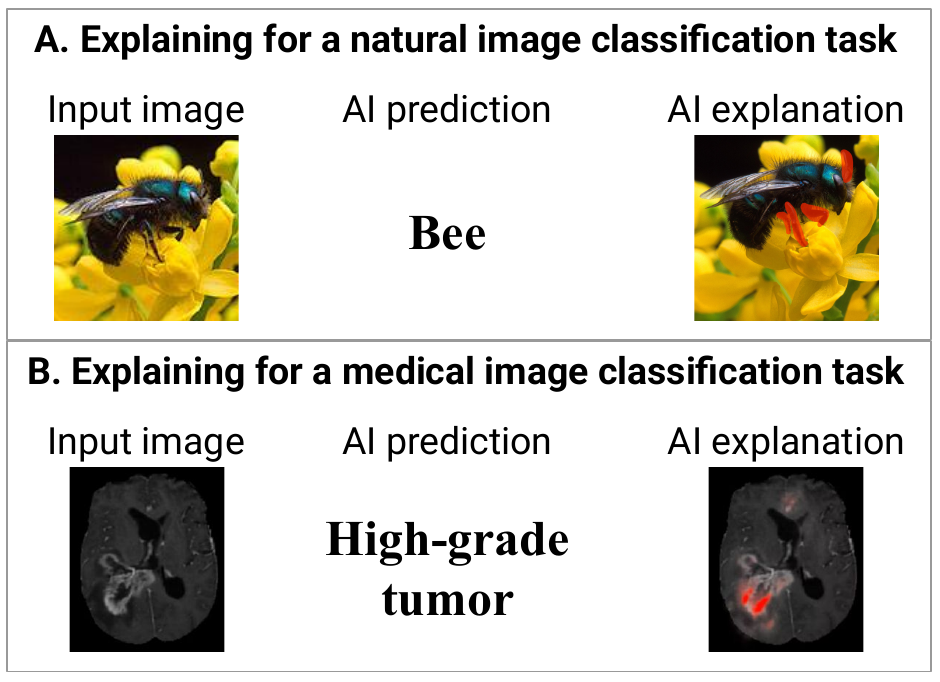}
    \caption{
    The feature attribution explanation on a natural image task of insect classification (panel A) and on a medical image tasks of brain tumor grading from MRI (panel B). The input image to an AI model and the AI prediction is indicated. The feature attribution explanation use a colored mask (in red) to highlight the important image pixels for the AI model’s prediction. The insect image is photographied by Jack Dykinga from USDA Agricultural Research Service (\href{https://www.ars.usda.gov/oc/images/photos/may00/k5400-1/}{image source link: https://www.ars.usda.gov/oc/images/photos/may00/k5400-1/}), and the brain MRI is from BraST 2020 dataset~\citep{Menze2015} with data ID BraTS20\_Training\_221. Both images are publicly available. The medical image only shows the MRI T1CE modality for illustrative purpose and other MRI modalities are not shown. 
    }
    \label{fig:feature_attri}
\end{figure}

For explanation forms that are user-friendly, one of the prevalent forms in the XAI literature is feature attribution\footnote{\label{fn:feat_attr}Feature attribution explanation is also called “heatmap”, “feature localization”, or “feature importance” in the XAI literature. In a systematic review on 312 papers that proposed an XAI technique \citep{10.1145/3583558}, among the 14 different forms of explanations, the top three frequent forms all belong to the feature attribution family, with 27\% feature importance (85/312), 21\% heatmap (67/312), and 21\% localization (65/312).}, shown in~\cref{fig:feature_attri}. In image recognition tasks, feature attribution explanations highlight the important regions for an AI prediction, using a color map overlaid on the input image~\citep{JIN2023102009}. This is like pointing at the salient part in the input image to users, which is intuitive to understand at first glance. However, when we presented such explanations to neurosurgeons in our clinical user study on an AI-assisted brain tumor grading task based on magnetic resonance imaging (MRI), doctors had a hard time understanding the meaning of the feature attribution explanation (\cref{fig:feature_attri}B)~\citep{JIN2024102751}. For example, in the study interview, doctors (N2 and N3) said~\citep{JIN2024102751}: 

\begin{displayquote} \textit{``What does that [image region highlighted by the feature attribution explanation] mean? Like hey, which part of my car gets my car moving? It should say press the accelerator. But yours would just show a dashboard of the car, and show that the accelerator had a little bit of red on it, this button had some red, that button had some red, but it's not an explanation. ... [Let's] go to an example, and you'll see, what about the red areas under MRI T1CE [T1 contrast-enhanced image modality]? Was it central necrosis [an important medical image feature for tumor grading]? But it couldn't be the central necrosis, because there's more central necrosis in the temporal lobe, and that area didn't get highlighted. So anyway, I don't know, it's just confusing.}

\textit{These [feature attribution] color maps were totally useless without text, without any context or explanation, like those details. The color maps were just pretty, but they didn't explain anything.''} (N3) \end{displayquote}

\begin{displayquote} \textit{``Though the color map is drawing your eyes to many different spots, but I feel like I didn't understand why my eyes were being driven to those spots, like why were these very specific components important? And I think that's where all my confusion was.''} (N2) \end{displayquote}

Similar critiques are also made by other doctors: In a user study with pathologists on AI-assisted pathology image biomarker quantification task~\citep{Evans2022}, doctors found feature attribution explanation “difficult to understand, distracting and/or confusing (P1, 2, 5),” “trivially understandable but not at all useful (P6),” “interesting but did not know what exactly to interpret from it (P4, 5),” and “it was unclear what was meant by ‘the most relevant pixels' (P6)." A highly-cited paper published in a premier journal of digital medicine commented that for feature attribution explanation, ``simply localising the region does not reveal exactly what it was in that area that the model considered useful.''~\citep{Ghassemi2021}

Why is the feature attribution explanation understandable in certain natural image explanation tasks (\cref{fig:feature_attri}A) but not in medical image tasks (\cref{fig:feature_attri}B)? The doctor’s comments above may imply the reason: the feature attribution explanation cannot provide contextual information for the important features. Explanation is a two-way communicative process. According to relevance theory in pragmatics~\citep{Relevance_Communication_and_Cognition}, understandability is achieved when the ostensive information provided by explainer is adequate for audiences to infer what the explainer wants to convey. Natural image features are easy to infer by pointing to their locations, whereas medical image features are more complex to describe, and merely localizing them is insufficient for doctors to infer what exactly the important features are. 
Despite its obvious drawback in real-world problems, the development of various XAI techniques for feature attribution explanation continues to be a predominant focus within XAI research~\citep{10.1145/3583558} (see~\cref{fn:feat_attr}), especially as explored through highly-abstracted toy problems, such as explaining daily objects in natural images. But real-world problems in application areas that have a significant demand for XAI are often marginalized.

\subsection{Case 3: XAI technical evaluation and the questionable criterion}

Like the real-world application of any technique, an XAI technique cannot claim to have certain functionalities, such as explainability, solely based on its design specifications or theoretical guarantees without test.
This fallacy is similar to stating that an AI predictive model can perform  accurate predictions without rigorously testing for its predictive functionality in given contexts. Failure to perform relevant assessments of the stated functionalities is not aligned with scientific or engineering standard, and is suspected of ethics washing \citep{aivodji2019fairwashing,10.1145/3375627.3375833,10.1007/978-3-031-47665-5_29}. However, the lack of scientific rigor in evaluation is commonly observed in XAI. In a systematic review of XAI evaluation methods~\citep{10.1145/3583558}, 33\% of the surveyed 312 papers that proposed an XAI technique and have published in prominent\footnote{According to~\citet{10.1145/3583558}’s definition using A$\star$ ranking in the CORE 2021 rankings. } AI conferences did not evaluate the XAI using rigorous methods including quantitative assessment and/or user study. 
Furthermore, the XAI evaluation methods themselves, if not being critically examined, can pose unethical risks, as demonstrated below. 

In our first XAI project~\citep{jin2021mapdoesfitall}, to evaluate the clinical utility of XAI for medical image classification tasks, we followed the evaluation conventions in the XAI literature. One of the evaluation criteria we used was plausibility. It assesses the reasonableness of an AI explanation by comparing it with human prior knowledge~\citep{jacovi-goldberg-2020-towards,10.1145/3583558}. Plausibility is widely used in XAI evaluation: in a systematic review of XAI evaluation methods, 34.3\% (62/181) of the surveyed papers used plausibility as the evaluation criterion, which is the top-chosen evaluation criterion among the twelve criteria~\citep{10.1145/3583558}. 

From a technical standpoint, using plausibility to judge the goodness of AI explanations seems to be totally legitimate: More plausible explanations indicate better AI predictions and more trust from users. However, when we showed AI explanations to neurosurgeons in our user study, the involvement of users’ perspectives began to reveal the problem of plausibility: When a doctor said he was satisfied with an AI explanation and could trust the corresponding AI prediction, we noticed that the AI prediction was actually \textit{incorrect}~\citep{JIN2023102684}. Explanations like this (wrong AI predictions with plausible explanations) can mislead and manipulate users to trust AI’s wrong predictions, and we named them misleading explanations in our later work that critically examines the evaluation criterion of plausibility~\citep{jin2024plausibilitysurprisinglyproblematicxai}. Our examination showed that using plausibility as the criterion to evaluate or optimize XAI techniques is both scientifically invalid and unethical: it is invalid because doing so encourages AI explanations to be plausible regardless of the underlying decision quality of the AI model; it is unethical because doing so increases the occurrence of misleading explanations~\citep{jin2024plausibilitysurprisinglyproblematicxai}.

Our examination also pointed out that an underlying cause for the XAI community to use plausibility as a prevalent criterion lies in the neglect of the adversarial nature of XAI~\citep{jin2024plausibilitysurprisinglyproblematicxai}. Performing a critical, adversarial role means to improve a system by identifying its potential problems, such as doctors improving patients' health by diagnosing diseases, teachers improving students' learning by spotting their weaknesses, or auditors improving the accountability of companies by auditing their potential errors and fraud. XAI has the adversarial nature to expose potential issues in AI models' predictive process and make them transparent and accountable. Although people love to see more plausible explanations from XAI, just like people love to receive positive reviews from doctors, teachers, or auditors, using plausibility as the XAI criterion is like incentivizing doctors, teachers, or auditors to produce as many positive reviews as possible, which subverts their original roles of acting adversarially. 
If the adversarial role of XAI is ignored, the legitimate situation of AI models being encouraged to \textit{learn} plausible features (analogous to patients pursuing more positive health reports) and the illegitimate situation of XAI techniques being encouraged to \textit{present} plausible features (analogous to doctors being incentivized to produce more positive reports) can easily be confused, which is the status quo.

\subsection{Case 4: XAI technical objective and the ``trade-off" framing}

In Case 3, despite the fact that using plausibility as the XAI criterion is unethical and can increase misleading explanations, an argument to support its use is that more plausible explanations can increase user trust and adoption of a high-performing AI, thus improving the overall task performance. Here, ethics is sacrificed (due to the manipulation of users' trust) for performance. This kind of trade-off narrative often frames the debate around explainability~\citep{Vilone2021}, although researchers have argued that the explainability vs. performance trade-off lacks scientific grounds~\citep{Rudin2019,Rudin2022}. The trade-off myth reflects the long-standing tension between the two cultures of explanation and prediction in statistics and machine learning ~\citep{breimanStatisticalModelingTwo2001,pmlr-v235-rudin24a}. 

But is the trade-off framing really inevitable? As we have mentioned in Case 3, explainability has the intrinsic adversarial role that is supposed to improve the AI model by exerting a counteracting force and exposing its problems. A simple analogy for this relationship is the brake and engine system in a car, in which the brake and engine “trade off” (or, using a better word and perspective, complement) each other to serve the same goal of driving safely and efficiently. Our following research experience shows that when the complementarity perspective replaces the trade-off perspective, different objectives can be discovered for XAI. 

\begin{figure}
    \centering
    \includegraphics[width=0.5\linewidth]{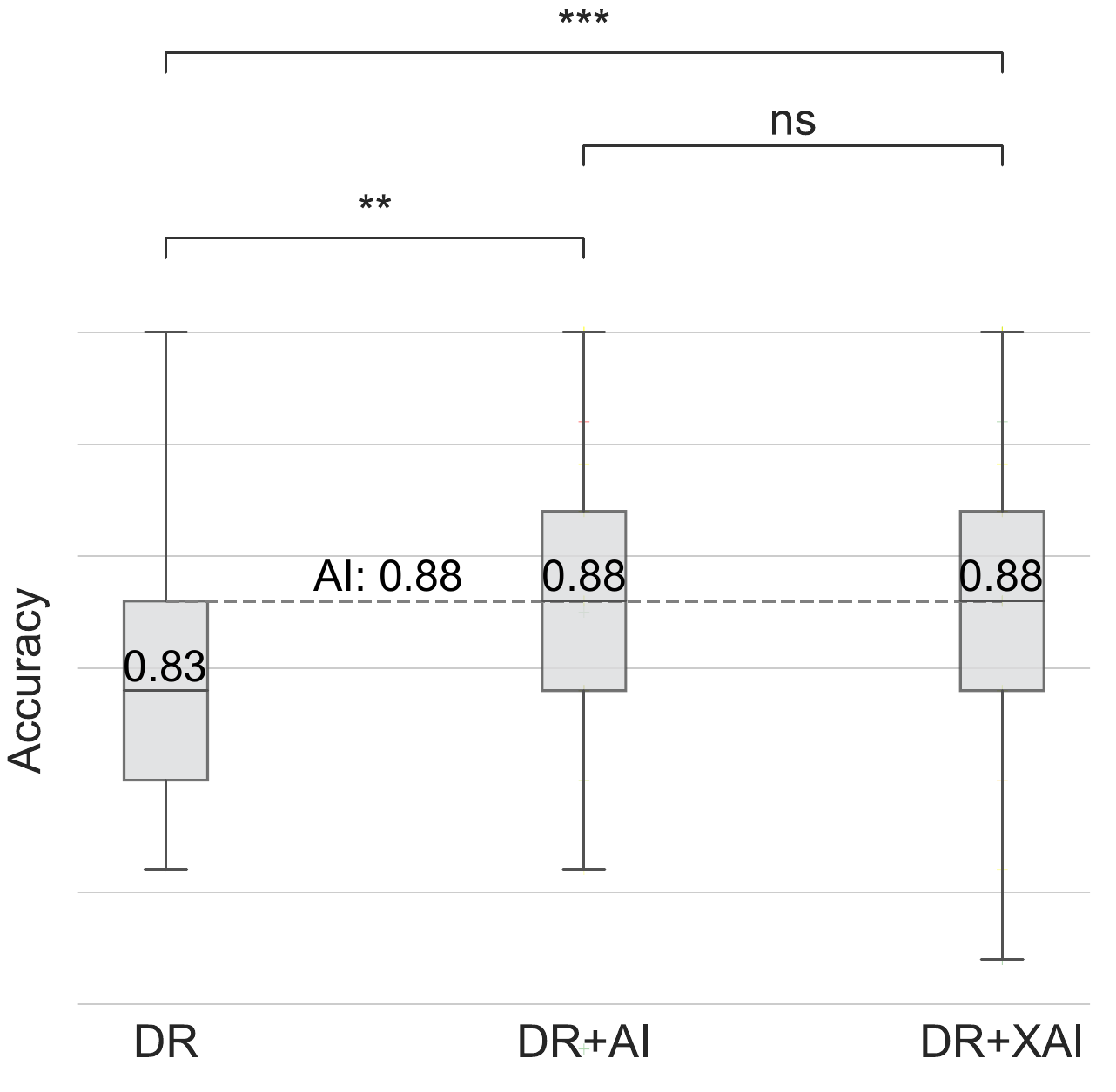}
    \caption{The neurosurgeon user study result in Case 4. It shows the 35 neurosurgeon participants’ medical image reading task accuracies in an experimental setting. The task was carried out in three conditions shown by the three box plots : when doctors performed the task alone (DR), with the prediction assistance of a black-box AI (DR+AI), and with the assistance of AI prediction and explanation (DR+XAI). Details about the study is in~\citet{JIN2024102751}. The result shows a repetitive pattern that can be seen in other similar AI-assisted studies, where with the help of a superior AI (the AI model’s accuracy 0.88 is visualized by the dotted line), human performance improved but not exceeding AI performance (*: statistically significant; ns: not significant). The same result may support distinct actions depending on different perspectives: it can either provide empirical evidence to support the replacement of humans with a black-box AI; or the empirical evidence can reveal flaws in the existing XAI approaches to improve both XAI and task performance.}
    \label{fig:doc_perf}
\end{figure}
In our neurosurgeon user study mentioned in Case 2, we wanted to know if the assistance of AI explanations can improve doctors' performance on grading brain tumor from MRI to be better than either doctors or AI performing the task alone~\citep{JIN2024102751}. This is termed as complementary task performance in the literature~\citep{10.1145/3411764.3445717, Zhang2020a,10.1145/3479552}. Our quantitative results showed that providing doctors with AI predictions from a black-box AI could improve doctors’ performance to be equivalent to AI’s performance, but the additional assistance from AI explanations did not further improve performance (\cref{fig:doc_perf}). 

At first glance, the results seemed to confirm the clinical utility of the assistance from black-box AI, but not from AI explanations. Then study results like this\footnote{Indeed, it is common to see similar study design and results like ours, in which the use of a higher performing black-box AI improved human performance, such as~\citet{Alufaisan_Marusich_Bakdash_Zhou_Kantarcioglu_2021, Ng2023,Yu2024}. But such study conclusions should be interpreted cautiously. The improvement in human performance with AI assistance may be explained away by the important latent factor of human over-reliance on AI, as we discuss later in this case.} can support the trade-off perspective that black-box AI is sufficient to improve performance, and explainability is useless and  redundant. However, we found such a performance improvement is not a good thing to celebrate, because it can be explained away by the factor of over-reliance (i.e., over-trust): our fine-grained analysis showed that doctors changed their decisions only when their initial decisions were different from AI’s suggestions (which can be correct or incorrect). This made doctors’ decisions more similar to AI’s regardless of AI decision correctness\footnote{This means the performance improvement is not because doctors accepted/rejected more of AI predictions when they should accept/reject (i.e., when AI predictions were correct/incorrect), but simply because doctors blindly accepted more suggestions from the high-performing AI regardless of AI correctness. }. This is a sign of over-reliance (or over-trust) on AI~\citep{Trust_in_Automation,bucinca2021trust, 7349687}, which undermines human autonomy, and can never achieve human-AI complementary performance\footnote{We proved theoretically that for classification tasks, with the assistance from a black-box AI, the human-AI team can never achieve complementary accuracy, see Lemma 1 Impossible Complementarity for Black-Box AI in~\citet{jin2024plausibilitysurprisinglyproblematicxai}. }. From a trade-off perspective, the over-reliance/over-trust phenomenon may be another reason to replace the human-AI team with a black-box AI~\citep{London2019,Durn2021}, since the involvement of human creates no gain in performance\footnote{It should be noted that this argument cannot withstand rigorous scrutiny, as it relies upon assumptions that ignore the complexity of real-world tasks: First, it assumes that the algorithmic performance tested in controlled, experimental settings can be generalizable to performance in real-world tasks. As ~\citet{frank_blind_2024} put it, ``The assumption is that how things behave in tightly controlled and manufactured environments should be our guide to how things behave in uncontrolled and unfabricated settings.” This is analogous to assuming that the test results of a new drug on mice can directly infer its effectiveness in patients. Second, it unreasonably assumes that the complex, open-ended, dynamic, and ambiguous real-world tasks can be reduced to and clearly defined in a closed form such that the outcome can be quantified by the one-dimensional metric of performance.}. And the technical objective should be focused on improving the performance of black-box AI models to surpass and replace humans. This is indeed the mainstream objective of AI in the technical community ~\citep{grace2024thousands, ai_imagination}. But we wanted to ask, by fixating on this single-minded objective, do we miss any important alternatives?

As we continued to analyze the fine-grained data on doctors' decision changes, we found that the reason for XAI failure is that AI explanations did not indicate to doctors when to rely on AI predictions and when not to, since after checking AI explanations, doctors had equal probabilities of changing their decisions correctly or incorrectly. From this finding, we later theoretically proved that as long as AI explanations can reliably indicate to users the uncertainties and potential errors of AI predictions, complementary human-AI performance can be achieved regardless of whether the AI's performance is superior or inferior to human’s\footnote{See Theorem 2 Conditions for XAI Complementarity and Corollary 3 in~\citet{jin2024plausibilitysurprisinglyproblematicxai}. This finding is related to the discussion on plausibility in Case 3.}~\citep{jin2024plausibilitysurprisinglyproblematicxai}. This finding provides a new technical objective for XAI to achieve both ethical purposes and better performance.

\section{Theoretical case analysis and the latent factor of power in AI ethics operationalization}\label{sec:analyze}

\begin{table}[!h]
    \centering
    \begin{tabular}{p{0.08\linewidth}p{0.38\linewidth}p{0.44\linewidth}}
    \toprule
\textbf{Case} & \textbf{Status quo} & \textbf{More ethical alternative} \\
\hline
\textbf{Case 1. XAI scope} & The target audiences of XAI in AI ethics are vague. & XAI in AI ethics should be oriented to non-technical audiences for algorithmic transparency and accountability. \\ \hline

\textbf{Case 2. XAI design}& XAI design is superficially understandable in abstracted toy problems that are distant from real-world problems. & XAI design should not marginalize the problem for non-technical audiences’ understandability of AI explanations.\\ \hline 

\textbf{Case 3. XAI evaluation} & The XAI community uses plausibility as the prevailing XAI evaluation criterion, which is scientifically invalid and unethical. & XAI evaluation should focus on the critical and adversarial role of AI explanations. \\ \hline
\textbf{Case 4. XAI objective}& Explainability is often framed as a trade-off with task performance. & 
By not framing explainability and performance as a trade-off, XAI that is aimed at revealing AI's weaknesses is theoretically shown to improve performance~\citep{jin2024plausibilitysurprisinglyproblematicxai}.\\ 
\bottomrule
    \end{tabular}
    \caption{Summary of the four cases described in~\cref{sec:case}. We contrast the status quo described in the cases, and the possible technical alternatives that are more aligned with the AI ethics principles.}
    \label{tab:case_summary}
\end{table}

\begin{figure}[!ht]
    \centering
    \includegraphics[width=0.6\linewidth]{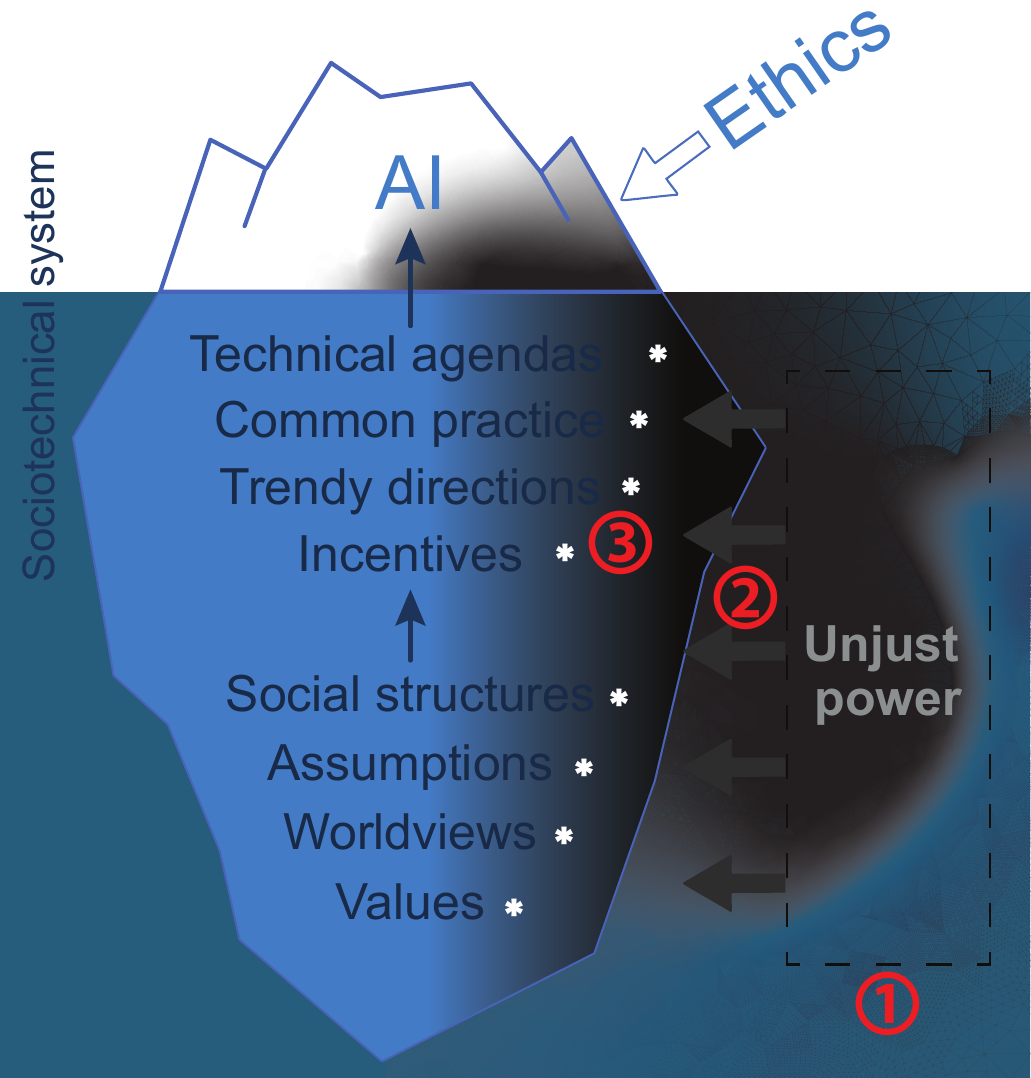}
    \caption{
    Illustration of the underlying mechanism on why it is difficult to implement ethics in AI in~\cref{sec:analyze}. It shows that AI is embedded in the sociotechnical system (iceberg surrounded by sea and air). AI (iceberg) and the social system (sea) coproduce each other. The technical aspect of AI (tip of the iceberg) is built upon the social and cultural aspects of the AI community (the underwater part of the iceberg). Power can be understood as particles that are diffusive in the sociotechnical system. It can be polluted by unjust power, illustrated by the black color. The red circled numbers highlight key points to intervene in the roots, which is the dysfunction in power mechanisms, detailed in \cref{sec:framework}: \Circled{1} by making power visible (\cref{sec:visible}); \Circled{2} by reflecting and resisting the framing of narratives and assumptions from unjust power (\cref{sec:frame}); and \Circled{3} by embedding ethical power (white particles) into the social aspect of AI (the underwater part of the iceberg, \cref{sec:method}).
    }
    \label{fig:iceberg}
\end{figure}

When we zoom out from the phenomena described in the four cases in~\cref{sec:case}, a repetitive pattern emerges: As summarized in~\cref{tab:case_summary}, in each case, we see that the current prevailing trends
in XAI development lifecycle --- from scope, design, evaluation, to objective --- all deviate from the original goals of explainability outlined by AI ethics principles. By describing these phenomena, we do not mean to deny the value of the XAI field nor good practices in XAI. Rather, the phenomena once again confirm the challenges of AI ethics implementation C1-C3 mentioned in the Introduction, and can be regarded as “symptoms” that are caused by their underlying “disease”. In this section, we aim to diagnose and understand the root causes of these phenomena: We first use the thematic analysis method~\citep{Braun2012} to analyze qualitative data from the four case descriptions
and identify common themes in the cases\footnote{The codes from the data analysis process are provided in~\cref{app:codes}.}. We then draw on theories from Science, Technology, and Society (STS), critical theory, and feminist epistemology to diagnose an underlying cause of the common themes and understand the mechanisms of the root cause. An overall of the disease  mechanism is illustrated in \cref{fig:iceberg}. 

\subsection{Overall preconditions: Sociotechnical system, social structure, and co-production}\label{sec:system}

The first clue we can get is that the phenomena described in the four cases are community-level prevalence, not a few exceptions from bad actors. This indicates that the root causes stem from the social system in which the technical community resides, rather than from individuals. To understand what social systems are and how they work, we need to abandon the individualistic perspective and utilize the knowledge and perspective of social systems. Sociology is the discipline to systematically study social systems~\citep{johnsonForestTreesSociology2014}\footnote{In~\cref{app:glossary}, we provide a Glossary of the new concepts introduced in this paper.}. Sociology views social system as
“[a]n interconnected collection of social structural relationships, ecological arrangements, cultural symbols, ideas, objects, and population dynamics and conditions”~\citep{johnsonForestTreesSociology2014} that cannot be simply reduced to its composing individuals. In the definition, social structure as “[t]he patterns of relationships and distributions” “characterize[s] the organization of a social system” ~\citep{johnsonForestTreesSociology2014}. 
This relational point of view rejects the individualism hypothesis that technological development occurs in a vacuum by lonely geniuses \citep{matthewman_technology_2011}. Instead, technologies are embedded in, shaped by, and inseparable from society. And the resulting technologies in turn shape society. This is termed the co-production of technology and society, and the intertwined system is named the sociotechnical system~\citep{matthewman_technology_2011,BoenigLiptsin2022}. 
Aided by these concepts, we can better describe our finding: the root causes that prevent the effective implementation of AI ethics reside in the sociotechnical system, characterized by the social structure.

\subsection{Power is an important latent factor in AI ethics implementation}

\begin{table}[h]
    \centering
    \begin{tabular}{p{0.12\linewidth}p{0.8\linewidth}}
    \toprule
\textbf{Case} & 
\textbf{Benefit whom?}
\\
\hline
\textbf{Case 1. XAI}\newline \textbf{scope} &  XAI is generally oriented towards technical audiences over non-technical audiences.\\ \hline

\textbf{Case 2. XAI}\newline \textbf{design}& XAI is largely designed to fulfill understandability in abstracted toy problems, problems that technical community is concerned about, over real-world problems where users mostly need XAI.\\ \hline 

\textbf{Case 3. XAI} \newline \textbf{evaluation} & 
a) The ubiquitous lack of scientific rigor in XAI evaluation while claiming to have the explainability functionality benefits the technical sector as a form of ethics washing~\citep{Floridi2019}. The consequences and harms of using inappropriately tested XAI are taken by users. 
\newline
b)  The use of plausibility as a prevelent criterion to evaluate XAI creates the effect of improving users’ “trust” in AI models’ wrong predictions, which benefits the technical sector to gain users’ trust but harms users. 
\\ \hline
\textbf{Case 4. XAI}\newline \textbf{objective}& 
XAI is often framed as an “inevitable” trade-off with performance. Performance typically represents the values and interests of the technical sector~\citep{birhane2022values}, while explainability embodies ethical values and user requirements. 
The unnecessary trade-off restricts the opportunity for users to advocate for their interests.
\\ 
\bottomrule
    \end{tabular}
    \caption{Description of the phenomena in the four cases to show that the tendencies in XAI scope, design, evaluation, and objective all favor the perspectives, interests, and benefits of the technical sector over users.}
    \label{tab:benefit-whom}
\end{table}

The next clue is in a consistent pattern in the four cases. 
All four cases deviate from ethical principles towards the same direction: the tendencies in XAI scope, design, evaluation, and objective all favor the perspectives, interests, and benefits of the technical sector over the users, as detailed in \cref{tab:benefit-whom}.
This pattern fulfills the definition of power, which is the capacity to have an effect, such as control and dominance, in spite of opposition (in our case, the opposition is the requirement for XAI from ethics principles)~\citep{johnsonForestTreesSociology2014}. Since the root cause is related to power and is a structural problem (as identified in~\cref{sec:system}), we can rule out powerful individuals and target power structure 
--- the distribution of power in a social system --- as the root cause. 
Specifically, \textbf{the root cause of the ineffective implementation of AI ethics is the dysfunctional power structure} in the sociotechnical system, which is lopsided to favor unethical power. It should be noted that we adopt a neutral definition of power, which does not imply that having power itself is negative or malign. Similarly, the root cause is not from having power, but from the dysfunction and imbalance of the power structure that causes the dominance of unchecked and unjust power\footnote{We use an analogy to emphasize the difference: the \textit{direct} cause of having a disease can be due to strong disease-causing factors (such as bacteria or oncogenes) or weak defensive mechanisms of the body (such as immune system), or both. In all these cases, the \textit{root} cause can be traced to the imbalance between the disease-causing factors and the body's defensive system. }. 
Next, by analyzing the common themes in the four cases, we aim to understand the specific mechanism of how power functions.

\subsection{How power works?}\label{sec:mechanism}
One mechanism for power to operate is by creating and shaping key configurations in social structures, cultures, assumptions, worldviews, values, and narratives in a social system~\citep{johnsonForestTreesSociology2014}. The system in turn “affects how we think, feel, and behave as participants … by laying out paths of least resistance in social situations. At any given moment, we could do an almost infinite number of things, but we typically do not realize this and see only a narrow range of possibilities. What the range looks like depends on the system we are in (ibid.).” This quote indicates that power and its enabling mechanism
largely operate in the “deep unconscious background” (ibid.), making them seem to be inevitable and “the natural order of things” (ibid.). Due to its hidden character and its distributed nature, the power structure as the root cause of ineffective implementation of AI ethics is often obscured in the technical community or even in the field of AI ethics. Next, to unveil the mechanisms of power structure, we describe the six themes (\cref{sec:structure}, \cref{sec:value}, \cref{sec:assumption}, and W1-3 in \cref{sec:worldview}) identified in our case analysis. These themes discuss the paths of least resistance in the social structure, values, worldview, narratives, and assumptions that are shaped by dominant power structures in the sociotechnical system.

\subsubsection{Social structures: Power shapes technical agendas, norms, and incentives}\label{sec:structure}

A way for power to shape social structures in a sociotechnical system is by shaping the paths of least resistance in research and development agendas, norms, common practices, and incentives in the technical community. 
Although technical practitioners or researchers can choose to work on any problem using any technical approach, this process does not happen in a vacuum but in a sociotechnical system. The technical community has regulations, rules, norms, standards, and conventions that lay the paths of least resistance for people to participate. The paths of least resistance outline which problems are worth investigating and which can be ignored, what kinds of technical choices are expected and what will face more challenges and be questioned. For example, in Case 1, researchers are expected and incentivized to use XAI to aid in paper publishing, while the real problem of using XAI to directly benefit users and challenge AI systems is marginalized. In Case 3, following the XAI evaluation conventions to use the plausibility criterion was our default approach as new comers to the XAI field (which was later critically examined by us to be wrong), because this approach is widely accepted in XAI. These paths create the least resistance because, by following them, people can gain incentives in the technical community (including recognitions, publications, fundings, or career advancements) much more easily than not following them~\citep{Rudin2013}. The mechanism of path of least resistance explains the community-level tendencies in \cref{sec:system}. The dominant power structures may exploit power imbalances and implicitly shape the paths of least resistance in the social structures that favor their own interests and benefits, thus perpetuating their dominance. 

These research and development agendas, technical norms, and common practices are shaped and justified by deeper social beliefs of worldviews, values, and assumptions, which are also influenced by power dynamics, as discussed in the following Sections \ref{sec:worldview}-\ref{sec:assumption}.

\subsubsection{The worldviews shaped by dominant power}\label{sec:worldview}
Our analysis reveals three epistemological-related themes (W1-W3) in the category of worldviews that are shaped by dominant power, as detailed below.

\vspace{5pt}
\noindent \textbf{W1. Presenting particular perspectives as general}

Both Case 1 and 2 use ambiguous users and their requirements to represent the ``general'' requirements of XAI regarding its scope and understandability. The ambiguous ``general'' users actually represent the perspectives and benefits of the dominant power in the sociotechnical system (see~\cref{tab:benefit-whom}). Meanwhile, all four cases show that the target users' needs and perspectives are largely overlooked. Case 2 shows that solving the ``general'' XAI problem on natural image explanation may not effectively address clinicians' specific XAI needs on understandability. 

This is a common epistemological fallacy that presents particular perspectives and interests (usually from the dominant power) as ``universal'' or ``general''~\citep{10.1145/3442188.3445943}.
The contemporary development of epistemology, especially with the contribution of feminist epistemology, has shown that all knowledge and knowers are always partial, socially situated, and associated with particular perspectives~\citep{Haraway1988,hardingWhoseScienceWhose1991,adamArtificialKnowingGender1998,CrasnowSharonL2024Feap,frank_blind_2024,Blackie2023}. Objectivity does not mean an impartial, disembodied,  disinterested, value-neutral perspective from a god's-eye view; instead, objectivity, “as a property of communities rather than individuals”, requires knowledge communities to “promote the epistemic and social aims of research, including minimizing bias”, and identify and scrutinize the encoded  values in knowledge and its production process~\citep{CrasnowSharonL2024Feap}. This means that there is no such a thing as a ``general” technology. Many pieces of evidence have shown that those depicted as technical ``standards” or ``universal” needs always come from the perspectives of the dominant power, often in the image of western, white, middle-class males~\citep{adamArtificialKnowingGender1998,Criado-Perez2019-qw, dignazioDataFeminism2020, CostanzaChock2023}. 

By presenting particular perspectives from dominant power as general, the perspectives and purposes of technology-for-power are disguised as technology-for-public or technology-for-users. As~\citet{Strathern2000-db} noted, ``power works most effectively when it is disguised.” Under this ``generality” worldview, power obscures its influences on technology and can remain unchallenged. Since the “generality” or “universality” narratives in technology are always associated with a hidden perspective and political agenda, instead of focusing on “how to develop technology that can benefit the masses?” we need to step back and ask, ‘“Generality” from whose perspective? For whom? For what purposes? And who could be harmed or ignored?’ Acknowledging the particular perspective in technology can improve our understanding of the problem, thus improve technological development. For example, Case 3 shows only when we switch perspective from technical people's to doctor's perspective can we identify the mistake in XAI evaluation criterion.

\vspace{5pt}
\noindent \textbf{W2. Substituting abstraction for real} 

W2 is closely related to W1, because presenting particular perspectives as general includes the process of abstraction and simplification.
In Case 2, we discussed the fallacy in XAI design agenda on understandability to focus too much on the decontextualized, highly-abstracted toy problems. As doctors’ feedbacks in Case 2 show and~\citet{pmlr-v235-rolnick24a} point out, ``ML algorithms designed in blue-sky, methods-focused research continue to fall short when used directly for applications.'' While abstraction is an essential means, it is fallacious to treat solutions to an abstracted problem as an end in AI development. 
Doing so implicitly substitutes the abstracted problems for the real, and fails to 
further understand, test, adapt solutions from abstraction to the real world. 
Being aware of the limitations of abstraction echoes with calls in the technical community for more work that grounds AI in the real-world, rather than abstract, problems and applications~\citep{pmlr-v235-rolnick24a, Rudin2013, 10.5555/3042573.3042809, ai_imagination}. 

The abstraction fallacy is not technically inevitable. 
Rather, it is a particular kind of worldview~\citep{frank_blind_2024} encouraged by the current dominant power in favor of its own interests. When the technical community gets too caught up in abstraction, the community can be easily manipulable, as technical people may be detached from their real users and actual problems, and become indifferent to the true aims and concrete social consequences of their work. As \cite{10.1145/15483.15484} noted, “[t]he actual problems on which one works - and which are so generously supported - are disguised and transformed until their representations are more fables, harmless, innocent, lovely fairy tales.” 
Another vivid example is from \citet{Kalluri2025}, in that they show how the systematic use of highly abstracted and obfuscating language in computer vision obscures the aim of developing AI for surveillance. 

\vspace{5pt}
\noindent \textbf{W3. Epistemic injustice and ignorance}

All four cases show a systematic downplay of users' roles and perspectives, while XAI is supposed to enable users to gain knowledge and understanding on AI predictions (\cref{tab:case_summary}). As we discussed in W1, because knowledge is always partial and socially situated, perspectives from diverse knowers are necessary to achieve more comprehensive knowledge. When power structures are imbalanced and generate systematic prejudices, one's capability as a knower can be suppressed and her epistemic resources can be denied due to her social position~\citep{Fricker2007,CrasnowSharonL2024Feap}. This is epistemic injustice, which is epistemically and socially harmful for both the oppressed knowers and the knowledge community. As shown in our cases, epistemic injustice harms both users and the AI community, as users lack effective XAI tools tailored to their needs to understand AI predictions~\citep{Symons2022} or can even be misled, and the XAI development is impeded due to the lack of involvement of users’ knowledge. 

Regarding the manifestation of epistemic injustice in AI, \citet{ethicalMISyn} presents two common technical assumptions that place one in a subordinate relationship, which are also seen in our XAI cases. The first one is ``placing humans in a subordinate position with respect to computational models in knowing and reasoning''~\citep{ethicalMISyn}. 
For example, in Case 4, AI is assumed to be more accurate and reliable than humans, if not now, in the foreseeable future~\citep{grace2024thousands}. So doctors have the obligation to rely on the high-performing AI's advices for patient benefits~\citep{Bjerring2020}. This assumption implicitly puts humans in an inferior position regarding our epistemic capabilities and contributions. It is untrue as it underestimates the complexity of real-world tasks and human capabilities in these tasks, and overlooks the intrinsic limits and fundamental dependency of AI on humans~\citep{ai_imagination}. 

The second one is placing some group of people in a subordinate position with respect to other groups regarding their epistemic capabilities and contributions. For example, in Case 1, the XAI scope prioritizes technical audiences' needs for explainability over non-technical audiences'. In Case 2, presenting particular perspectives as general is also an example that marginalizes perspectives from the less powerful, as analyzed in W1. 

While the less powerful suffer from epistemic injustice, the behavior of the powerful is ignorance, which has been systematically studied in epistemology~\citep{Tuana2006,Lassen2025}. Ignoring knowledge that can potentially challenge the dominant power is an effective way to perpetuate power imbalance~\citep{Crasnow2024ignorance}. For example, in Case 1, the oversight role of XAI for non-technical audiences is ignored in XAI scope. In Case 3, the critical and adversarial role of XAI is ignored in XAI evaluation. 

\subsubsection{Social values: Power encodes its favoring values in technology }\label{sec:value}
The analysis in~\cref{sec:worldview} W1 shows that because technologies are always socially situated, and always encode knowers’ perspectives and values on what are important, technology is always value-laden \citep{CrasnowSharonL2024Feap,verbeekMoralizingTechnologyUnderstanding2011a, ChinYee2019}. 
\citet{birhane2022values} identify top values encoded in AI research, including performance and efficiency; and these values support the centralization of power. 
Case 4 shows that the value of performance or efficiency, if prioritized, can exclude pluralistic values:
With the trade-off perspective that prioritizes the value of performance, even though ethical value is emphasized in the beginning, which utilizes a user-centered approach that centralizes doctors’ role in clinical decision-making, we can undergo a slippery slope to gradually exclude the more ethical options in order to pursue performance, first by providing plausible but misleading explanations, then by regarding explainability as redundant and removable, and eventually by replacing the human-AI team with black-box AI. 
The invasive nature of the value of performance can explain why it is difficult to implement AI ethics principles. 
In contrast, at the end of Case 4, we show that with the complementary perspective, by prioritizing the inclusive value of ethics, 
it is possible to develop XAI to achieve both explainability and high performance. 
Case 4 shows that power can encode its favoring values in AI, which can lead to the framing of distinct AI objectives and the resulting AI techniques accordingly.

\subsubsection{Power can shape assumptions in prevalent framing and narratives for its own favor}\label{sec:assumption}

In addition to shaping the worldviews and values that underpin AI technical practices, power can also shape the background assumptions in the prevalent framings and narratives in AI technical practices. Doing so can eliminate potentially unfavorable alternatives at the very early stage, making the included options and their underlying values and assumptions seemingly natural and inevitable. 

For example, Case 4 presents a prevalent framing of the performance vs. explainability trade-off in XAI. A similar and common narrative in AI ethics is that AI ethics/regulations may impede technical development~\citep{schaakeTechCoupHow2024}. These framings and narratives prematurely exclude the obvious facts that explainability and performance, or AI ethics and technological development, are for the same goal and do not frame a trade-off at all, as we have addressed in W3 on ignorance, and the brake and engine analogy showing the non-linear, adversarial role of ethics in Case 4. Furthermore, with different perspectives, values, and assumptions, the same empirical evidence in Case 4 (the higher performance of black-box AI and the ineffectiveness of XAI, \cref{fig:doc_perf}) can be interpreted differently to frame distinct technical narratives and objectives: the assumption to prioritize ethics (the complementarity perspective) interprets the result as an opportunity to adjust the current objective of XAI; while the assumption to prioritize performance (the trade-off perspective) prematurely dismisses this opportunity, and attributes the failure of XAI to its intrinsic nature (rather than to the wrong objective), thus denying the value of explainability.

\subsection{Summary}
We summarize our case analysis in this section, including our diagnosis and the understanding of the problem mechanism.

\textbf{The diagnosis. } 
We diagnose the root cause of the ineffective implementation of AI ethics in AI practice being the dysfunctional power structure in the sociotechnical system, which is imbalanced to favor unethical power. 

\textbf{The mechanism of power. } 
Power influences the paths of least resistance that shape how people participate in a sociotechnical system. 
Namely, power can influence the sociotechnical structures, perspectives, worldviews, values, assumptions, and narratives underpinning the agendas, visions, norms, technical standards, common practices, and evaluation criteria of AI technical practices, in a way that favors the power's own values and interests and perpetuates its dominance. Ethics is meant to represent power from the public and civil society, especially for those who are marginalized, vulnerable, and negatively impacted by AI. The ineffectiveness in AI ethics implementation means that the power structures of the sociotechnical system where AI technologies reside is dysfunctional and imbalanced that the current dominant power underpinning AI does not represent the power of the public and the marginalized communities, and the power from ethics and the public cannot countervail the dominant power in AI.

\section{Making power explicable: power-sensitive interventional recommendations  for AI ethics implementation}\label{sec:framework}

Our case analysis in \cref{sec:analyze} shows AI ethics operationalization is not a linear process but involves complex sociotechnical interactions and dynamics, in which power is an important latent factor. To facilitate the effective implementation of AI ethics, we propose the power-sensitive interventional recommendations. Our recommendations consist of three aspects based on understanding of the power mechanism analyzed in \cref{sec:analyze} on the sociotechnical, framing, and epistemological aspects of ethics and power (\cref{fig:iceberg}), detailed in the next three subsections. 
A checklist summary of the interventional recommendations is in \cref{tab:recommand}.

\subsection{Ethics is political: Making power explicable and checked}\label{sec:visible}

Our case study shows that the underlying power dominating XAI technical development is unjust, as it prioritizes a narrow technical agenda while marginalizing the original ethical agenda for users and public interests. This finding aligns with many pieces of prior evidence to show that the technical agenda can be hijacked by those who have capital and/or political power to serve their own values and interests~\citep{Ahmed2023,Ochigame2022,10.1145/3531146.3533194,birhane2022values,katzArtificialWhitenessPolitics2020,Burrell2024,Ali_Dick_Dillon_Jones_Penn_Staley_2023,wigginsHowDataHappened2023,chunDiscriminatingDataCorrelation2021,Gebru2024,Penn_2023,10.1145/3375627.3375813,10.1145/15483.15484}. For technical people in the AI community, this indicates for us that if we don’t realize the power- and value-laden nature of technology, or pretend AI or AI ethics is apolitical, we can easily be manipulated, and the technical agendas can be suboptimal or distorted, as shown in the XAI cases. 

Power, as the capacity to have an effect, is diffusive and inevitable in the sociotechnical system (\cref{fig:iceberg}). 
In this sense, power is analogous to bacteria, which are normal existences in the living environment. So the goal is not to create a power-free environment for AI and eliminate the influence of power to AI (analogous to creating a bacteria-free living environment), which is unrealistic and impossible. The interventional goal is to restore the normal function in the power structure (analogous to maintaining a healthy immune system) to be able to coexist and maintain checks and balances with power (bacteria). 
As we mentioned earlier, if “power works most effectively when it is disguised”~\citep{Strathern2000-db}, then the primary goal is to make power visible and explicable, such that its potential influence and biases can be scrutinized and checked to (re)direct power for just purposes. 

Ethics is an important countervailing force to keep dominant power in check, and its agenda should include making power explicable. This is analogous to developing a strong immune system for the AI community. 
Checking power means ethical and responsible AI should include power analysis in its technical research and development processes. In major and minor design choices in AI technical development, such as problem identification, formulation, data collection, model design, training, evaluation, deployment, and monitoring, technical people should answer these questions~\citep{dignazioDataFeminism2020}: For whom are we making such a decision? For what purposes? In which contexts? And who can be put at risk? Are we aware that discourses such as “general” audiences or “universal” needs can present particular perspectives as “general” that disguises unjust power?
Faithfully answering these questions can help us to be reflective and explicit about what the major underlying power is that guides such a choice, and how different choices would potentially shift power dynamics. 
Such inquiries should also be included in the ethics review and analysis routine.
Qualitative and quantitative methods can be explored to identify power imbalances in AI technical practices. Prior works such as~\citet{birhane2022values,10.1145/3531146.3533157,Kalluri2025} provide methods to identify the underlying power and values in AI research papers and patents, which can be useful resources to refer to. It should be noted that the process of power analysis can face the same challenges C1-C3 as the process in ethics operationalization. To prevent the same problem of ineffective implementation of power analysis, the methodologies and implementation of power analysis should be reflective of their own power dynamics and aware of the potential penetration of unjust dominant power in the operationalization process. 

Along with power analysis, the AI community needs to unite with the public and civil society, especially with communities that are marginalized or negatively impacted by AI, such that we can take collective actions to gain power and use ethics to check and balance the current dominant power underlying AI. In the next two subsections, we present two strategies on power check based on our case analysis.

\subsection{Ethics as the brake to complement the engine for justice: Reflecting and changing the framing of AI and ethics}\label{sec:frame}

\begin{table}[!h]
    \centering
    \begin{tabular}{p{0.35\linewidth}| p{0.55\linewidth}}
    \toprule
\textbf{The current framings of AI and AI ethics underpinning by unjust power} & \textbf{Reframing AI and AI ethics underpinning by just power} \\
\hline
The aims of AI technologies are for technical innovation, efficiency, and performance~\citep{birhane2022values}. & Using the car analogy, these aims are for the engine, not for the car. AI's developmental goals are not to build good engines only, but to build good cars that includes the building of strong engines and equally strong brakes. The developmental goals of AI technologies are to improve scientific and social benefits as well as to minimize the risks and harms. \\ \hline
Ethics is apolitical. & The operationalization of ethics should be power-sensitive, not circumvent the root causes of unethical issues in AI. \\ \hline
The implementation of ethics may impede AI innovation and reduce how much AI produces benefits to society. & AI benefits are not power- and value-neutral, and can easily be hijacked by unjust power that creates benefits for a few but misery for others. AI innovation with weak ethics in place is like a car having strong engines and weak brakes. The adversarial role of ethics (the brake) is to complement AI innovations (the engine) for the same scientific and social aims (the aims of a car). \\ \hline 
The implementation of AI ethics is to develop techniques on AI fairness, explainability, privacy, and security. & These AI ethics approaches are technological fix~\citep{TechnologicalFix2024} that may relieve symptoms of AI harms but not treat the disease, which is related to sociotechnical factors of social and power structures, values, worldviews, etc. The implementation of AI ethics should tackle both the disease and the symptoms by embedding power-sensitive ethics approaches into AI technical practices of all AI subfields and tasks. \\ 
\bottomrule
    \end{tabular}
    \caption{The comparison of different framings of AI and AI ethics according to distinct underpinning power. The second column shows the reframing of the narratives of AI and AI ethics based on just power.}
    \label{tab:reframe}
\end{table}

For AI ethics to check power, it needs to have teeth~\citep{doi:10.1177/2053951720942541} and act adversarially. As Case 3 demonstrated, the XAI evaluation criterion is incorrect just because the adversarial role of XAI is ignored. We use the word “adversarial” to denote the role of improving AI through a non-linear, antagonistic process by identifying and pruning AI's potential risks, harms, and weaknesses, which is analogous to doctors/teachers helping patients/students by identifying their diseases/weaknesses. A competitive, exclusive perspective, however, may only focus on the antagonistic behavior, and prematurely excludes the original cooperative role of AI ethics. This is like falsely framing the brake as a tradeoff with the engine, and ignoring the fact that they complement each other for the same goal of driving safely and efficiently. As Case 4 shows, the false binary between AI ethics and AI benefits (such as innovation or efficiency) is not uncommon in the current framing of AI ethics implementation, and it is shaped by the perspective and values of the dominant unjust power. Therefore, a strategy to check power is to be critical of the narrative and power that frame AI and AI ethics. We reframe several key narratives about AI and AI ethics based on just power in~\cref{tab:reframe}.

\subsection{Ethics as technical methods: Critical and limitation analyses for scientific and ethical rigor}\label{sec:method}

\begin{table}[!h]
    \centering
    \begin{tabular}{p{0.3\linewidth}| p{0.6\linewidth}}
    \toprule
\textbf{Research direction} & \textbf{Example} \\
\hline
Critically examining assumptions, norms, conventions, and standards in AI technical practices & XAI evaluation criterion ~\citep{jin2024plausibilitysurprisinglyproblematicxai}; Evaluation criteria for computer vision~\citep{Reinke2024}; Dataset~\citep{9423393, NEURIPS_DATASETS_AND_BENCHMARKS2021_3b8a6142} \\ \hline
Critically examining the AI technique practice paradigm & The general AI vision~\citep{blilihamelin2025stoptreatingaginorthstar,ai_imagination,10.1145/15483.15484, 10.5555/3042573.3042809,Rudin2013,koch2024protoscienceepistemicmonoculturebenchmarking}; Values in AI research~\citep{birhane2022values, BIRHANE2021100205}; XAI paradigm~\citep{Rudin2019}; Goals of computer vision~\citep{Kalluri2025}; Reproducibility crisis of AI~\citep{Kapoor2023}; Pseudoscience in AI~\citep{Andrews2024}\\ \hline
Identifying and understanding the limitations of AI techniques & Limits of AI modeling~\citep{10302088}; Limits of generative AI~\citep{Shumailov2024}; Epistemic limitations of generative AI in science~\citep{Messeri2024}\\ \hline
 Explicitly encoding ethical values and perspectives as technical standards for an AI task or subfield
   & Ethical framework for the medical image synthesis task covering its whole lifecycle, including problem formulation, design, evaluation, limitation analysis, peer review, and public oversight ~\citep{ethicalMISyn}\\ 
\bottomrule
    \end{tabular}
    \caption{List of research directions and their examples of uniting ethics and scientific practice for AI's scientific rigor and social justice. }
    \label{tab:method}
\end{table}

Our analysis of epistemic injustice shows that ethics has the potential to improve our ways of knowing, and the pursuits of truth (i.e., science) and goodness (i.e., ethics) can complement each other: prior research has shown that unethical conduct leads to unscientific AI research~\citep{Andrews2024}; our Case 3 also shows that rigorous scientific evaluation is the gatekeeper for the successful implementation of AI ethics techniques. The adversarial role of ethics for power check aligns with the self-critique and self-correction mechanisms in scientific conduct. Therefore, the two efforts of implementing AI ethics and improving the scientific rigor of AI research can unite for responsible AI. For example, the real-world harms of AI can be due to the lack of scientific rigor in AI assessment, which fails to account for the inevitable discrepancies between the expected performance of AI in a controlled, predictable environment and its actual deployment in the open, complex, and unpredictable reality~\citep{10302088}. At the operational level, the perspective and values of AI ethics can be explicitly encoded as methodologies in technical practices to improve the rigor and robustness of AI techniques when embedded in complex social contexts. Just as the dominant power encodes its values and perspectives in the technical practices in the case study, we can encode ethical values in the same manner to reshape common technical practices. 
In~\cref{tab:method}, we list several research advances and examples to inspire follow-up works to continue exploring this much-needed research direction.

Among the efforts of encoding ethics as scientific methods, we highlight two approaches: critical examination and limitation analysis. They both join the adversarial role of ethics and the self-critique and correction mechanism in scientific research to focus on the negative aspects in AI techniques, such as weaknesses, risks, harms, and questionable underlying assumptions. Doing so can help maintain health for AI development, similar to the adversarial role of doctors or teachers in identifying problems. 
It can also help mitigate the current biased view of AI capabilities that creates epistemic injustice as we analyzed in~\cref{sec:worldview}, which tends to overestimate AI capabilities, underestimate the complexity of real-world tasks, and downplay the contribution of human labor to AI~\citep{neda_surrogate_2019}. Using the car analogy again, the strong momentum of AI development (the engine) should be equipped with equivalently strong ethics (the brake). Paying proportional attention to the negative aspects of AI gives ethics the teeth and power to reject AI when AI is inappropriate~\citep{Benjamin2016,10.1145/3630107}. 

Specifically, \textbf{critical examination} is to encourage more reflections on the taken-for-granted practices and their assumptions in AI. If unjust power exerts its influence via the behavior of uncritically following conventions, then critical reflection and examination are the immune system to defend against it. We list several examples of critical examination in~\cref{tab:method}. Critical examination can become institutionalized as an AI research topic or direction, for example, by becoming one of the research focuses in the Call For Paper agendas of the influential AI conferences and publishing venues.

\textbf{Limitation analysis} is to identify, understand, and evaluate the scopes, limitations, weaknesses, risks, harms, failure modes of an AI technique, and the discrepancies between algorithmic abstraction and real-world complexity. Compared to more mature disciplines such as biomedicine and engineering that prioritize safety over efficacy in assessment standards, the AI field can borrow experience from these disciplines,  and investigate in equivalent amount of resources and attention to limitation analysis as the analysis on algorithmic performance and efficiency. 
Limitation analysis can have three aspects: 1) developing new methods to qualitatively and quantitatively assessing the scope, weaknesses, failure modes, and potential risks and warnings of an AI technique. 2) The acknowledgment of the design-specific assumptions and limitations in problem formulation, model design, and evaluation methodology that may be mitigated in future work. 3) The acknowledgment of the intrinsic limits of an AI technique and its evaluation method that cannot be overcome with technological progress, such as the intrinsic limits of abstraction and ``generality'' in AI modeling (see W1, W2 in~\cref{sec:worldview}), the discrepancies between training data and the real-world data, the intrinsic gap between the closed-world settings of the assessment method and the open-world settings when AI technique is deployed.  
All aspects can be institutionalized as standard evaluation and reporting methods. For example, prior research~\citet{ethicalMISyn} has demonstrated the embedding of the three aspects of limitation analysis in the peer review process. 

\section{Conclusion}

In this work, based on the symptoms presented in the explainable AI case study, we draw from social theories and epistemology to diagnose a possible root cause preventing the effective implementation of AI ethics, which is the dysfunctional and imbalanced power structures in the sociotechnical system of AI that favor unjust power. We then provide the interventional recommendations to tackle the roots, including making power explicable and checked in AI ethics, reframing AI and AI ethics to avoid unacknowledged and unjust assumptions, and conducting critical and limitation analyses.

Our empirical evidence of the case study is limited by our research experience in XAI, and future works are needed to verify or modify our diagnosis with other empirical evidence. 
Our work aims to diagnose the ethical issues in AI through the power lens. We acknowledge that complex problems like this can be understood and tackled from pluralistic perspectives. Other different perspectives are out of the focus of this work, and can be explored in other and future works. 
We hope our diagnosis and the interventional recommendations can be a useful input for the AI community and civil society's iteration and trial and error in the real world, to improve the ethics of AI technologies and enable AI to directly achieve its intended scientific and social aims.

\part*{Appendix}
\appendix

\section{Glossary}\label{app:glossary}

\textbf{AI practices} or \textbf{AI technical practices}. In this paper, we use the phrases to denote the day-to-day practices and activities of AI technical professionals in the whole pipeline of AI research, development, deployment and maintenance, etc.

\textbf{Adversarial role}. In this paper, we use the word “adversarial” to denote the role of improving a system through a non-linear, antagonistic process by identifying and pruning the system's potential risks, harms, and weaknesses, which is analogous to doctors/teachers helping patients/students by identifying their diseases/weaknesses.

\textbf{Epistemic injustice}. It is when one's capability as a knower is suppressed and her epistemic resources is denied due to her social position~\citep{Fricker2007,CrasnowSharonL2024Feap}. 

\textbf{Epistemology}. “Epistemology is the study of knowledge, including the difference between knowledge and other cognitive states.”~\citep{CrasnowSharonL2024Feap}

\textbf{Ethics washing}. Ethics washing, or ethics bluewashing in~\citet{Floridi2019}, is ``the malpractice of making unsubstantiated or misleading claims about, or implementing superficial measures in favour of, the ethical values and benefits of digital processes, products, services, or other solutions in order to appear more digitally ethical than one is.''~\citep{Floridi2019}

\textbf{Ethics shopping}. Ethics shopping, or digital ethics shopping in~\citet{Floridi2019}, is ``the malpractice of choosing, adapting, or revising (“mixing and matching”) ethical principles, guidelines, codes, frameworks, or other similar standards (especially but not only in the ethics of AI), from a variety of available offers, in order to retrofit some pre-existing behaviours (choices, processes, strategies, etc.), and hence justify them a posteriori, instead of implementing or improving new behaviours by benchmarking them against public, ethical standards.''~\citep{Floridi2019}

\textbf{Co-production}. Co-production is a concept in Science, Technology and Society (STS) that “view[s] data science and the social world to be mutually constitutive. ... Co-production says that our ideas about what the world is—as a matter of fact—and what the world ought to be—as a matter of social values—are formed together (Jasanoff 2004). Instead of privileging either the role of social forces or technological forces as explanations for the way that the world is, the co-productionist approach considers the symmetrical interplay between social and technical forces in weaving the fabric of the world. These co-evolving forces can be observed and discovered and become sites of deliberate work and transformation.

Co-production gives us insight into how the prevailing conditions of life and authorized visions of the future come to define technological design and deployment, and how these technological designs shape what is deemed possible and desirable for society. As a result, we see that technology is not alone responsible for determining the state of human oppression today across axes of race, gender, and class nor does it alone determine the possibility for social justice. And yet technology plays a constitutive role in defining the specific ways in which empowered and marginalized peoples relate to one another, forge their identities and shape their lives. In other words, we are able to see how certain forms of life and technologies cohere together, and how other forms of life are written off as superfluous, unvalued, or impossible. In contrast to these coherent ways of life, alternative visions and resistances to the existing order are thrown starkly into view. By providing insight into the differently possible, the co-productionist approach supports people to make social change.”~\citep{BoenigLiptsin2022}

\textbf{Ignorance}. ``Ignorance in the realm of science is typically depicted as a gap in knowledge: something that we do not (yet) know.''~\citep{Tuana2006} `Social theories of ignorance, however, focus on the ways in which social and cultural forces construct and maintain ignorance. They examine how such forces work to bring about not only a lack of knowledge, but make it difficult, if not impossible to know certain things (Smithson 2008: 214–215). Thus, the focus here will be on ignorance not as a “neglectful” epistemic practice, but as “a substantive epistemic practice in itself” (Alcoff 2007: 39).'~\citep{CrasnowSharonL2024Feap} \citet{Tuana2006} provides a taxonomy of ignorance, including:

1. What we don’t know, but don’t care to know. 

2. What we don’t know that we don’t know.

3. What they don’t want us to know. 

4. What they don’t know and don’t want to know. 

5. What we cannot know.

\textbf{Individualism}. Individualism is ``[a] way of thinking based on the idea that everything that happens in social life results solely from the thoughts and feelings of individuals without reference to their participation in social systems.''~\citep{johnsonForestTreesSociology2014}

\textbf{Justice and injustice}. “Justice and injustice concern primarily an evaluation of how the institutions of a society work together to produce outcomes that support or minimize the threat of domination, and support or minimize everyone's opportunities to develop and exercise capacities for living a good life as they define it.”~\citep{young2003political}

\textbf{Narratives}. “The concept of \textit{narratives}  refers to prominently circulated or widely held beliefs about why the world is as it is, what changes are possible and worthwhile, and what should be done to achieve a desired future. Narratives are profoundly important in human life for individual and collective sense-making and accounting (Ricøeur 1988; Butler 2005; Meretoja 2018) and are constructed around any and every facet of existence. In the context of the data science workflow, we are particularly concerned with how narratives garner support for technological and scientific interventions, and how, in turn, technologies underwrite visions of futures worth wanting and working toward (Jasanoff and Kim 2015).”~\citep{BoenigLiptsin2022}

\textbf{Objectivity}. In this paper, we adopt the understanding of objectivity from feminist epistemology, in which it rejects an impartial, disembodied,  disinterested, value-neutral perspective from a god's-eye view. Instead, objectivity, “as a property of communities rather than individuals”, requires knowledge communities to “promote the epistemic and social aims of research, including minimizing bias”, and identify and scrutinize the encoded  values in knowledge and its production process~\citep{CrasnowSharonL2024Feap}. 

\textbf{Path of least resistance}. “In a social system, the path of least resistance in a particular situation consists of whatever behavior or appearance is expected of participants depending on their position in that system.”~\citep{johnsonForestTreesSociology2014}

\textbf{Politics and Political}. ``Politics is the set of activities that are associated with making decisions in groups, or other forms of power relations among individuals, such as the distribution of status or resources.''~\citep{wiki_politics}

\textbf{Power}. 
Power is a multifaceted concept and rejects a single best definition~\citep{Waelen2022}. To emphasize the oppressive as well as the emancipatory potential of power, we synthesize multiple viewpoints and provide a working definition of power as ``[t]he ability to have an effect, including the assertion of control and dominance, in spite of opposition”~\citep{johnsonForestTreesSociology2014}. “Power is a socially situated capacity to control others' actions.”~\citep{Fricker2007} 
And “social power is a capacity we have as social agents to influence how things go in the social world.”~\citep{Fricker2007}  
Power “describe[s] the current configuration of structural privilege and structural oppression, in which some groups experience unearned advantages --- because various systems have been designed by people like them and work for people them --- and other groups experience systematic disadvantages --- because those same systems were not designed by them or with people like them in mind”~\citep{dignazioDataFeminism2020}. 

\textbf{Power structure}. Power structure is ``[t]he distribution of power in a social system.''~\citep{johnsonForestTreesSociology2014}

\textbf{Situated knowledge}. Proposed by~\citet{Haraway1988}, situated knowledge is “one of the key features of feminist epistemology. ... To claim that knowledge is situated is to claim that knowledge is for and by socially located knowers and so both the production of knowledge and the knowledge produced is always local. The idea that knowledge is situated involves not only a rejection of the abstract knower but also a rejection of knowledge as timeless and universal – true everywhere and always – a Platonic conception of knowledge.”~\citep{CrasnowSharonL2024Feap}

\textbf{Social structure}. Social structure is ``[t]he patterns of relationships and distributions that characterize the organization of a social system. Relationships connect various elements of a system (such as social statuses) to one another and to the system itself. Distributions include valued resources and rewards, such as power and income, and the distribution of people among social statuses. Structure can also refer to relationships and distributions among systems.''~\citep{johnsonForestTreesSociology2014}

\textbf{Social system}. Social system is ``[a]n interconnected collection of social structural relationships, ecological arrangements, cultural symbols, ideas, objects, and population dynamics and conditions that combine to form a whole. Complex systems comprise smaller systems that are related to one another and the larger system through cultural, structural, ecological, and population arrangements and dynamics."~\citep{johnsonForestTreesSociology2014}

\textbf{Sociotechnical system}. Sociotechnical system is “[t]he concept of sociotechnical systems concerns the ways in which people and technologies are irrevocably entangled across different scales such that human agency is intertwined with technological devices, practices, and infrastructures.”~\citep{BoenigLiptsin2022} 

\textbf{Structures}. “[S]tructures refer to the relation of social positions that condition the opportunities and life prospects of the persons located in those positions. This positioning occurs because of the way that actions and interactions reinforce the rules and resources available for other actions and interactions involving people in other structural positions. The unintended consequences of the confluence of many actions often produce and reinforce opportunities and constraints, and these often make their mark on the physical conditions of future actions, as well as on the habits and expectations of actors. This mutually reinforcing process means that the positional relations and the way they condition individual lives are difficult to change."~\citep{young2003political}

\textbf{Value}. Value is ``[a]n idea about relative worth, goodness, or desirability used to make choices among different alternatives. In a patriarchy, for example, maleness is culturally valued above femaleness, and being in control is valued above not being in control.''~\citep{johnsonForestTreesSociology2014}

\textbf{Worldview}. Worldview is ``[t]he collection of interconnected beliefs, values, attitudes, images, stories, and memories out of which a sense of reality is constructed and maintained in a social system and in the minds of individuals who participate in it.''~\citep{johnsonForestTreesSociology2014}
\section{Codes from qualitative data analysis}\label{app:codes}

We list the codes used in the qualitative data analysis in~\cref{sec:analyze}. The co-author WJ conducted the data analysis process. We aggregated the codes to form the six themes described in~\cref{sec:mechanism}, as listed in~\cref{tab:aggregate}.

\begin{table}[h]
    \centering
        \caption{Six themes aggregated from the codes}
    \begin{tabular}{p{0.45\linewidth}|p{0.48\linewidth}}
    \toprule
\textbf{Theme} & \textbf{Codes} \\ \hline
Social structure (\cref{sec:structure})&
$\bullet$ Systematic issues 

$\bullet$ Benefits whom? Who are at risk?
\\ \hline

Dominant worldview: objectivity, presenting partial perspective as general (W1 in \cref{sec:worldview}) & 
$\bullet$ Lack user's perspectives

$\bullet$ Presenting particular interests as general \\ \hline
Dominant worldview: Taking the abstraction as the real (W2 in \cref{sec:worldview}) & 
$\bullet$ Taking the abstraction as the real

$\bullet$ AI not grounded in real users and real-world applications \\ \hline
Dominant worldview: Epistemic injustice and ignorance (W3 in \cref{sec:worldview}) & 
$\bullet$ Epistemic injustice for users 

$\bullet$ Ignorance \\ \hline
Dominant values (\cref{sec:value})& 
$\bullet$ Technology is value-laden 

$\bullet$ Values encoded in technology \\ \hline
Dominant narrative (\cref{sec:assumption}) & 
$\bullet$ Problem framing and narrative \\ 
\bottomrule
    \end{tabular}
    \label{tab:aggregate}
\end{table}
\FloatBarrier

    \begin{longtable}{p{0.25\linewidth}|p{0.68\linewidth}}
   \caption{Codes in Case 1}    \label{tab:code1}
   \endfirsthead
    \endhead
    \hline
\textbf{Code in Case 1} & \textbf{Coded texts} \\ \hline
Benefits whom? Who are at risk? &  ``most of us as AI researchers are building explanatory agents for ourselves, rather than for the intended users.'' 

“[m]any participants who used XAI methods in practice stated that their primary motivation for using XAI was to aid in publishing their papers rather than improving user comprehension.'' 

“the majority of deployments are not for end users affected by the model but rather for machine learning engineers, who use explainability to debug the model itself.”

“explanations primarily serve internal stakeholders rather than external ones.”
\\ \hline
Lack user's perspectives & 
``We find that current approaches to using XAI in AI4SG create a severe incentive misalignment, devaluing the needs of users in favor of publication practices in machine learning that often prioritize model performance over user interpretability.” 

“[m]uch of XAI research build explanations for those \textit{who are already AI experts}. Overall, little regard is given to non-AI-expert stakeholders who need AI explanations, and have different goals for those explanations. This produces a large gap between technical experts’ version of XAI, and what non-technical audiences expect of XAI.”
\\ \hline
Systematic issues & 
Paragraph 2 on the description of XAI for technical audiences rather than nontechnical audience, and XAI for publication rather than users. Those are not due to individual choices but are guided by systematic factors. \\ \hline

Presenting particular interests as general & “[m]uch of XAI research build explanations for those \textit{who are already AI experts}. Overall, little regard is given to non-AI-expert stakeholders who need AI explanations, and have different goals for those explanations. This produces a large gap between technical experts’ version of XAI, and what non-technical audiences expect of XAI.”\\ \hline
Epistemic injustice for users: Create epistemic barriers for user's understanding and transparency; Users lack epistemic resources to understand AI  & “There is thus a gap between explainability in practice and the goal of transparency, since explanations primarily serve internal stakeholders rather than external ones.”\\ \hline
Ignorance, what we don’t know, but don’t care to know & Paragraph 2\\ \hline
AI not grounded in real users and real-world applications & 
``most of us as AI researchers are building explanatory agents for ourselves, rather than for the intended users.'' 

“little regard is given to non-AI-expert stakeholders who need AI explanations, and have different goals for those explanations.”  \\ \hline
    \end{longtable}
\FloatBarrier

\begin{longtable}{p{0.25\linewidth}|p{0.68\linewidth}}
        \caption{Codes in Case 2} \label{tab:code2}
    \endfirsthead
    \endhead
    \hline
\textbf{Code in Case 2} & \textbf{Coded texts} \\ \hline

Ignorance, what we don’t know, but don’t care to know & The doctors’ quotes showing the problems in feather attribution explanation, and the statistics showing the prevalence of the feature attribution explanation in XAI \\ \hline

Taking the abstraction as the real & 
`Like explainability, understandability is always coupled with its audience to whom and in which context an AI explanation is understandable. An XAI technique designed for the technical or “general” audiences and purposes may not be understandable for non-technical audiences in specific contexts.'

The last paragraph showing the abstracted toy problem of natural image explanation lacks concrete contextual information that can sufficiently describe the image features for the real-world problem of medical image explanation \\ \hline

Presenting particular interests as general & 
“An XAI technique designed for the technical or “general” audiences and purposes may not be understandable for non-technical audiences in specific contexts”

The last paragraph 
\\ \hline

AI not grounded in real users and real-world applications & The last paragraph \\ \hline

Epistemic injustice for users: Create epistemic barriers for user's understanding and transparency; Users lack epistemic resources to understand AI & The doctors’ quotes showing they didn’t understand the feature attribution explanation \\  \hline
\end{longtable}
\FloatBarrier

    \begin{longtable}{p{0.25\linewidth}|p{0.68\linewidth}}
   \caption{Codes in Case 3}    \label{tab:code3}
   \endfirsthead
    \endhead
    \hline
\textbf{Code in Case 3} & \textbf{Coded texts} \\ \hline

Benefits whom? Who are at risk? & “Failure to perform relevant assessments of the stated functionalities is not aligned with scientific or engineering standard, and is suspected of ethics washing \citep{aivodji2019fairwashing,10.1145/3375627.3375833,10.1007/978-3-031-47665-5_29}. However, the lack of scientific rigor in evaluation is commonly observed in XAI. In a systematic review of XAI evaluation methods~\citep{10.1145/3583558}, 33\% of the surveyed 312 papers that proposed an XAI technique and have published in prominent AI conferences did not evaluate the XAI using rigorous methods including quantitative assessment and/or user study.” 

“more plausible explanations indicate better AI predictions and more trust from users. …Explanations like this (wrong AI predictions with plausible explanations) can mislead and manipulate users to trust AI’s wrong predictions, and we named them misleading explanations.”
\\ \hline

Lack user's perspectives & “when we showed AI explanations to neurosurgeons in our user study, the involvement of users’ perspectives began to reveal the problem of plausibility”  \\ \hline

Ignorance, what we don’t know, but don’t care to know & “Our examination also pointed out that an underlying cause for the XAI community to use plausibility as a prevalent criterion lies in the neglect of the adversarial nature of XAI~\citep{jin2024plausibilitysurprisinglyproblematicxai}.” \\ \hline

Technology is value-laden & The choice of whether to use plausibility as an XAI criterion or not reflects two different values and perspectives of 1) focusing on technical sectors’ benefits and interests only, and 2) aligning technical sectors’ benefits and interests with users’ benefits and interests \\ \hline

Systematic issues & 
“However, the lack of scientific rigor in evaluation is commonly observed in XAI. In a systematic review of XAI evaluation methods~\citep{10.1145/3583558}, 33\% of the surveyed 312 papers that proposed an XAI technique and have published in prominent AI conferences did not evaluate the XAI using rigorous methods including quantitative assessment and/or user study.”

“In our first XAI project~\citep{jin2021mapdoesfitall}, to evaluate the clinical utility of XAI for medical image classification tasks, we followed the evaluation conventions in the XAI literature. One of the evaluation criteria we used was plausibility. It assesses the reasonableness of an AI explanation by comparing it with human prior knowledge~\citep{jacovi-goldberg-2020-towards,10.1145/3583558}. Plausibility is widely used in XAI evaluation: in a systematic review of XAI evaluation methods, 34.3\% (62/181) of the surveyed papers used plausibility as the evaluation criterion, which is the top-chosen evaluation criterion among the twelve criteria~\citep{10.1145/3583558}.” \\ \hline
\end{longtable}
\FloatBarrier

\begin{longtable}{p{0.25\linewidth}|p{0.68\linewidth}}
   \caption{Codes in Case 4}    \label{tab:code4}
    \endfirsthead
\endhead
    \hline
\textbf{Code in Case 4} & \textbf{Coded texts} \\ \hline

Values encoded in technology & Trade-off perspective: prioritizing the value of performance

Complementarity perspective: prioritizing the value of ethics

“The same result may support distinct actions depending on different perspectives”

\\ \hline

Problem framing and narrative & 
“But is the trade-off framing really inevitable? … Our following research experience shows that when the complementarity perspective replaces the trade-off perspective, different objectives can be discovered for XAI.”

“The same result may support distinct actions depending on different perspectives: it can either provide empirical evidence to support the replacement of humans with a black-box AI; or the empirical evidence can reveal flaws in the existing XAI approaches to improve both XAI and task performance.”
\\ \hline

Lack user's perspectives & “This is a sign of over-reliance (or over-trust) on AI~\citep{Trust_in_Automation,bucinca2021trust, 7349687}, which undermines human autonomy, and can never achieve human-AI complementary performance”\\ \hline

Ignorance & “But is the trade-off framing really inevitable? As we have mentioned in Case 3, explainability has the intrinsic adversarial role that is supposed to improve the AI model by exerting a counteracting force and exposing its problems. A simple analogy for this relationship is the brake and engine system in a car, in which the brake and engine “trade off” (or, using a better word and perspective, complement) each other to serve the same goal of driving safely and efficiently.” \\ \hline

Taking the abstraction as the real & “It should be noted that this argument cannot withstand rigorous scrutiny, as it relies upon assumptions that ignore the complexity of real-world tasks: First, it assumes that the algorithmic performance tested in controlled, experimental settings can be generalizable to performance in real-world tasks. As ~\citet{frank_blind_2024} put it, “The assumption is that how things behave in tightly controlled and manufactured environments should be our guide to how things behave in uncontrolled and unfabricated settings.” This is analogous to assuming that the test results of a new drug on mice can directly infer its effectiveness in patients. Second, it unreasonably assumes that the complex, open-ended, dynamic, and ambiguous real-world tasks can be reduced to and clearly defined in a closed form such that the outcome can be quantified by the one-dimensional metric of performance.”\\ \hline

Epistemic injustice & “From a trade-off perspective, the over-reliance/over-trust phenomenon may be another reason to replace the human-AI team with a black-box AI~\citep{London2019,Durn2021}, since the involvement of human creates no gain in performance”\\ \hline

\end{longtable}
\FloatBarrier

\small
\bibliographystyle{rusnat}
\bibliography{xai_power}
\end{document}